\documentclass[%
reprint,
 amsmath,amssymb,
 aps,
]{revtex4-1}

\usepackage{graphicx}
\usepackage{dcolumn}
\usepackage{bm}
\usepackage{mathrsfs,amsmath}

\begin{document}

\preprint{APS/123-QED}

\title{Towards uncovering the structure of power fluctuations of wind farms}

\author{Huiwen Liu$^1$, Yaqing Jin$^2$,  Nicolas Tobin$^2$, Leonardo P. Chamorro$^{*2,3,4}$}

\affiliation{1. Visiting student. College of Water Conservancy and Hydropower Engineering, Hohai University, Nanjing, China}
\affiliation{2. Department of Mechanical Science and Engineering, University of Illinois, Urbana, IL, 61801}
\affiliation{3. Department of Aerospace Engineering, University of Illinois, Urbana, IL, 61801}
\affiliation{4. Department of Civil and Environmental Engineering, University of Illinois, Urbana, IL, 61801}

\date{\today}

\begin{abstract}
The structure of the turbulence-driven power fluctuations in a wind farm is fundamentally described
from basic concepts. A derived tuning-free model, supported with experiments, reveals the underlying spectral content of the power fluctuations of a wind farm. It contains two power-law trends and oscillations in the relatively low- and high-frequency ranges. The former is mostly due to the turbulent interaction between the flow and the turbine properties; whereas the latter is due to the advection between turbine pairs. The spectral wind-farm scale power fluctuations $\Phi_P$ exhibits a power-law decay proportional to $f^{-5/3-2}$ in the region corresponding to the turbulence inertial subrange and at relatively large scales, $\Phi_P\sim f^{-2}$. Due to the advection and turbulent diffusion of large-scale structures, a spectral oscillation exists with the product of a sinusoidal behavior and an exponential decay in the frequency domain.

\end{abstract}

\pacs{Valid PACS appear here}
\maketitle

\section{Introduction}
 Wind is  a mainstream source of electricity, and will play a leading role in achieving climate goals. Fundamental understanding on the relation between turbulence and wind turbines is key to improve reliability, predictability, and integration of wind farms into electrical grids. 

Turbulence plays a dominant role in the structure of a wind farm's power output. In particular,  turbulence intensity ($I_u$) is closely associated with power fluctuations \cite{Rosen96}, fatigue accumulation, \cite{Frandsen99} as well as forces and bending moments \cite{Van08}. High turbulence can increase the mixing of wakes and thus alter the mean velocity and turbulence levels near downwind turbines  \citep{Ozbay12}. Simple analytical models are widely used to characterize wakes, including mean velocity (e.g., \citep{Jensen83,Bastankhah14,Pena14,Chu14}) and $I_u$ (e.g., \citep{Hassan93,Xie14,Niayifar15,Gocmen16,Vermeer03,Sorensen11,Sanderse11}). Particular emphasis has been placed on the structure of the velocity fluctuations. \citet{Crespo96} proposed a spectrum model for the evolution of wind-turbine wakes. \citet{Chamorro12} pointed out that wind turbines act as an 'active filter' of flow by modulating the large and small scales. \citet{Howard15a} and \citet{Chamorro15a} noted that the flow structures developing from upstream bluff bodies may leave strong signature on the fluctuations and spectrum of the power output of wind and hydrokinetic turbines. Recently, \citet{Jin16} showed the distinctive effect of background flow in the intermediate field and the increasing growth rate of the integral scale with turbulence. 

Substantial effort has  been placed on turbulence effects in wind farms.  \citet{Sorensen02} proposed a model for the interaction between wind farms and power systems based on the turbulence spectrum. Milan \textit{et al}. \citep{Milan13} suggested that for large time scales, the power  fluctuations of wind farms can be considered to follow adiabatic wind dynamics with a similar $f^{-5/3}$ spectral behavior. However, recent work by \citet{Bandi17} has shown that the effect of geographical smoothing on aggregate wind power outputs indicate an asymptotic limit of $f^{-7/3}$ for disperse wind farms. A similar observation was made by Apt for time scales ranging from 30 s to 2.6 days \citep{Apt07}. Chamorro \textit{et al}. \citep{Chamorro15b} showed three regions in the spectral domain defined by dynamical aspects of the flow and its interaction with the turbine. The power output appears insensitive to turbulence in the high-frequency region, where the turbulent scales are smaller than the rotor. In the intermediate region, with length scales up to those on the order of the atmospheric boundary layer thickness (ABL), the spectral content of the power fluctuations $\Phi_P$ and flow $\Phi_u$ exhibit a relationship characterized by a transfer function $G(f)\propto f^{-2}$. In the low-frequency range, very large-scale motions (with sizes on the order of the ABL and larger) directly impart their spectral characteristics onto the power output, and approach the $f^{-5/3}$ behavior observed by other authors. More recently, Tobin \textit{et al}. \citep{Tobin15b} proposed a tuning-free model for $G(f)$ to predict power fluctuations of single turbines, which includes the modulation of the turbulence structure and the mechanical characteristics of the wind turbine. Mur-Adama \textit{et al}. \citep{Mur07} proposed that the sum of the frequency components of a single turbine approximates the wind farm output. However, it has since been observed that inter-turbine correlations have a marked effect on spectral structure, shown in field data by \citet{Calif13}, the large-eddy simulations of \citet{Stevens14}. and porous disk experiments by  \citet{Bossuyt17}.

Despite these efforts, a gap still remains in the quantitative description of the power fluctuations of wind farms as a function of the incoming turbulence, which is a building block for improving their efficiency and life span.  This work aims to fill this gap by deriving wind-farm power fluctuations from first principles supported with experiments.

 \section{Experimental Setup}

Wind tunnel experiments with two aligned wind farm models were performed to quantify the bulk power fluctuations and to test the developed model both for single turbine and wind-farm-scale power fluctuations. It is worth stressing that the model is scale-agnostic, and is able to predict the power structure of these model turbines and those at field scale 1 kW and 2.5 MW, as demonstrated by Tobin \textit{et al}. \citep{Tobin15b}.

\begin{figure}
 	\begin{center}
 		\includegraphics[width=1\linewidth]{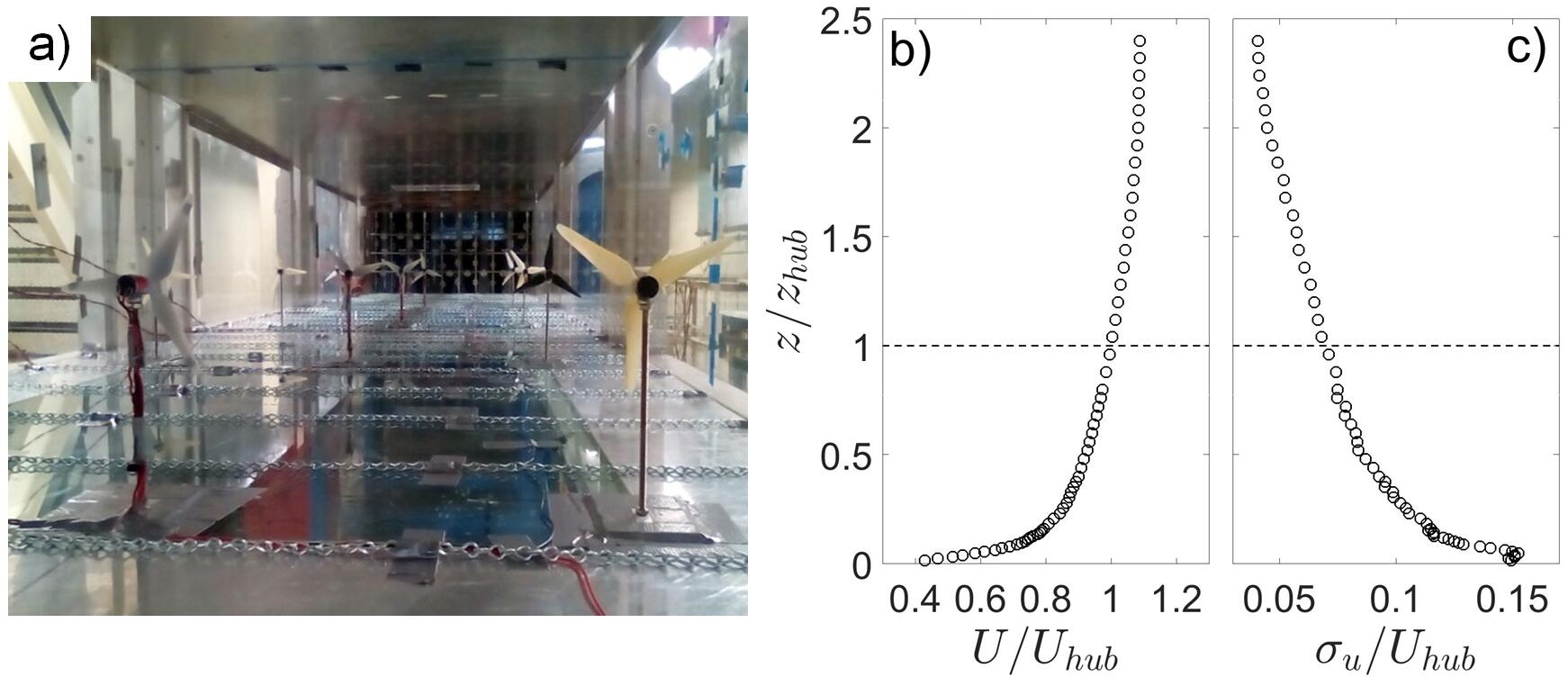}
\caption{a) Photograph of the test section pointing upwind; b) Mean velocity $U/U_{hub}$; c) turbulence intensity $I_u=\sigma_u/U_{hub}$. Horizontal lines indicate the turbine-hub height.}%
 	\label{photo}
 	\end{center}
 	\end{figure}

Model wind farms were operated in the Talbot wind tunnel under nearly zero pressure gradient (fig. \ref{photo}). The test section is 6.1 m long, 0.914 m wide, and 0.45 m high \citep{Adrian00}. An active turbulence generator \citep{Jin16}  created a realistic turbulent shear flow containing an inertial subrange spanning two decades. Roughness consisting of 5 mm chains every 0.2 m \citep{Ohya01,Chamorro09} was also placed along the test section to develop a turbulent boundary layer (see Fig. \ref{photo}b-c). The turbines are based on a reference model from Sandia National Laboratory \citep{Johnson13,Shiu12}. The rotors have a diameter $d_T$ = 120 mm and hub height $z_{hub}$ = 125 mm \citep{Tobin15a}. A Precision Micro-drives 112-001 Micro Core 12 mm was used as the loading system, with a rated power $P_0 \sim$ 1 W. Additional characteristics quantities of the turbine can be found in Tobin et al \cite{Tobin17}. 

The distance $\Delta_x$ between turbines was $S_x=\Delta_x/d_T = 7$ and 10 in the flow direction, whereas both configurations has $S_y=\Delta_y/d_T = 2.5$ in the transverse direction. This resulted in 6$\times$3 and 5$\times$3 turbine arrays, where power measurements were performed on the central turbines. The experiments were conducted with an incoming hub-height velocity of $U_{hub}=$ 9.71 ms$^{-1}$ giving a Reynolds number $Re=U_{hub} d_T/\nu \approx 7.56\times10^4$. The turbines operated at a tip-speed ratio of $\lambda=\omega d_T/(2U_{hub})\approx4.9$, where $\omega$ is the angular velocity of the rotor. The measured power coefficient for the turbine is $C_p\approx 0.08$. This low value is due to the inefficiency of the generator (around $\sim$20\% at the rotational speeds during the experiments) and not indicative of poor aerodynamic performance \cite{Johnson13, Tobin17}. The estimated thrust coefficient $C_T\approx0.5$. The boundary layer had a thickness of $\delta/z_{hub} \approx 2.4$ and friction velocity $u_\ast \approx 0.46$ ms$^{-1}$. 

Flow data were obtained from a high-resolution hotwire anemometer with the height adjusted by a bidirectional slide positioning system mounted at the top of the wind tunnel. The sensor voltage signatures were sampled at 10 kHz for a measurement period of 90 s when characterizing the boundary layer. 
Hotwire measurements were also taken in the upwind vicinity of each turbine to get the local incoming flow at a frequency of 20 kHz for a period of 120 s. A Measurement Computing USB-1608HS data acquisition system was connected to the generators to collect the instantaneous turbine voltages. Output power was measured at 100 kHz for a period of 120 s and inferred from the voltage and the terminal resistance ($2\Omega$) of the generator.

\section{Results and discussion}
\subsection{Power fluctuations of turbines in wind farms}

To characterize the structure of the power fluctuations of wind farms, it is informative to first describe the fluctuations of individual turbines. Tobin \textit{et al}. \citep{Tobin15b} proposed an analytical model that accounts for the underlying physical filtering process performed by a wind turbine in response to incoming turbulence. Based on the energy balance of the turbine's rotor,
\begin{equation}
	dE_{rot}/dt = -P + 0.5C_P \rho A u^3\left( t\right) 
	\label{rotorenergy}
\end{equation}
where $\rho$ is the air density, $A$ is the swept area of the rotor, $E_{rot}=Pt_i$ is the mechanical energy and $P$ is the power. Here, $t_i = I\omega/ 2\tau$ is the inertial timescale that depends on the properties and operation of the turbine, $I$ is the moment of inertia of the rotor and $\tau$ is the electric torque.  Solving equation \ref{rotorenergy} with a Green's function results in the following transfer function $\hat{G}\left(f \right)$ for $\Phi_P$:
\begin{equation}
	{| \hat{G}\left( f\right)|}^2={t_i}^2/[1+4\pi^2f^2{t_i}^2] 
	\label{transferfunction}
\end{equation}
where  $\Phi_P=\hat{G}(f)\Phi_u$, with $\Phi_u$ representing the velocity spectrum of the incoming flow. As $f \to 0$, $\hat{G}\left(f \right) \to {t_i}^2=const$. This flat response at low frequencies is consistent with observations where $\Phi_P$ appears to be proportional to $\Phi_u$. However, as $f$ increases, $\hat{G}\left(f \right) \to f^{-2}$. A similar phenomenon occurs in the case of wind arrays, which is explored as follows.

\begin{figure}
	[h!]
	\centering
	\includegraphics[width=1\linewidth]{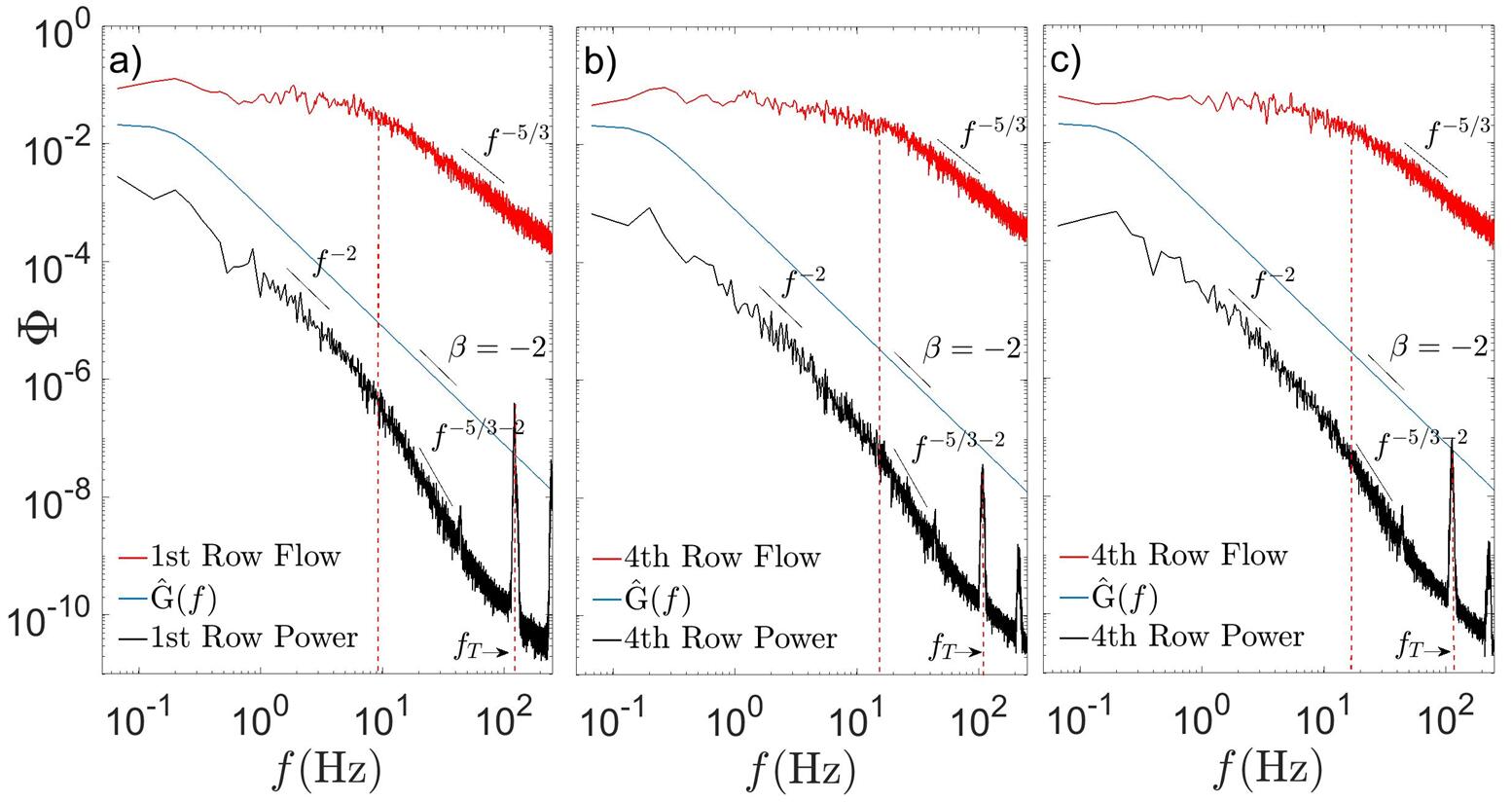}
	\caption{Spectra of incoming turbulence ($red$), $\hat{G}(f)$ ($blue$) and turbine output power ($black$) of a) 1st and b) 4th row, $S_x = 7$; c) 4th row, $S_x = 10$.}
	\label{power_flow_spec}
\end{figure}

The distributions of $\Phi_P$, and $\Phi_u$ directly upwind of the rotors, for the central turbines at the 1st and 4th rows in the two setups is shown in Figure \ref{power_flow_spec}; the function $\hat{G}\left(f \right)$ is included as a reference. There, the peaks correspond to the turbine rotational frequency $f_T$ and  harmonics. The distinctive modulation of the flow structure and the turbine power via $\hat{G}(f)$ is made clear in this Figure. In particular, the power fluctuations of the turbines in the 4th row in the two setups also exhibit regions with spectral decay of $f^{-2}$ and $f^{-2-5/3}$, but the location where they occur varies. Note that the beginning of the $f^{-2-5/3}$ region is shifted to a higher frequency in the 4th row. This is due to the difference in the wind farm layout, which modulates the structure and evolution of the turbulence inside the wind farm with respect to that of the incoming flow. 

\begin{figure}
	[h!]
	\centering
	\includegraphics[width=0.8\linewidth]{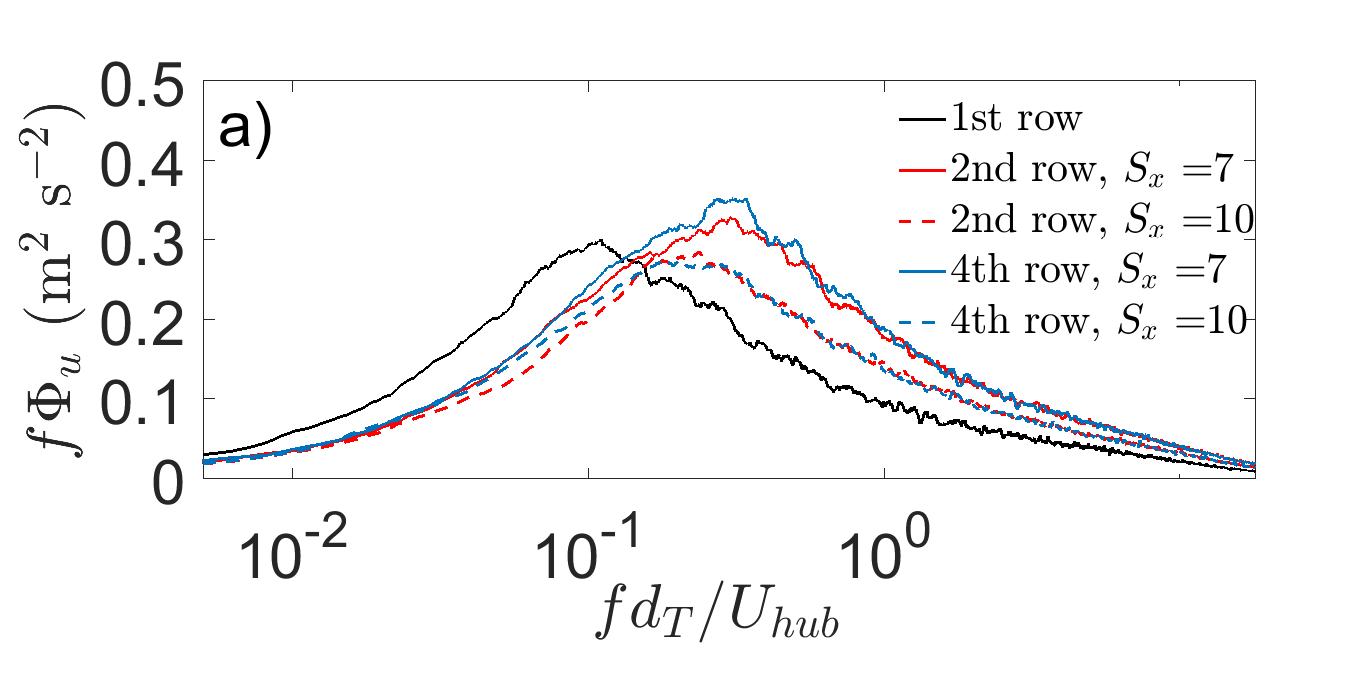}
	\includegraphics[width=0.8\linewidth]{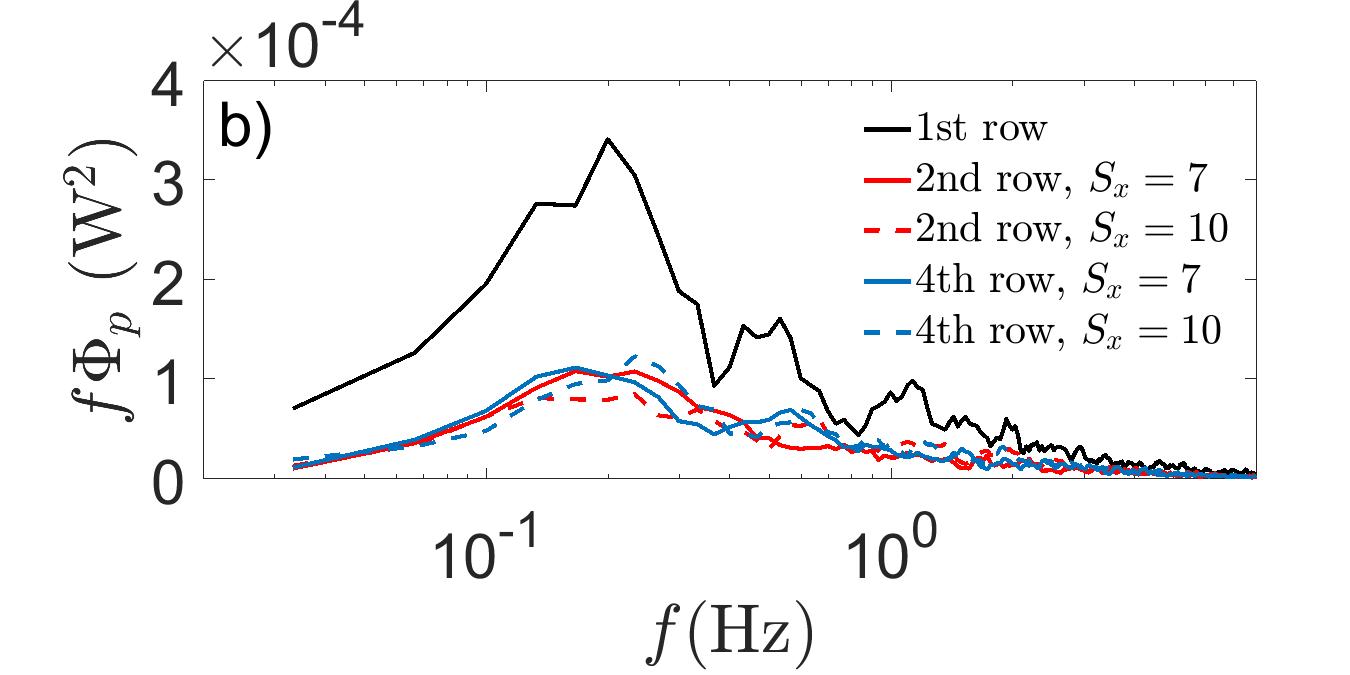}
	\caption{Pre-multiplied power spectra of a) the local incoming velocity and b) the power output of the 1st, 2nd and 4th rows with $S_x = 7$ and 10 (solid and dotted lines).}
	\label{power_pre_spec}
\end{figure}

Further, the pre-multiplied spectra of the local incoming velocity $f\Phi_u$ and power output $f\Phi_P$ for the 1st, 2nd and 4th rows in the two setups are shown together in Figure \ref{power_pre_spec}. The representative turbulent scale of the incoming flow at hub height is larger than that of those within the wind farm due to the modulation of the wind turbines; this effect is reduced with increased turbine spacing. Compared with the inner rows, the power fluctuations of the 1st row are more energetic across all scales. The differences between the inner rows is much smaller, as flow velocity, $I_u$ and integral length scale do not vary substantially. 
\vspace{-3mm}

\subsection{Wind farm power fluctuations}
\begin{figure}
	[h!]
	\centering
	\hspace{-3.2mm}
	\includegraphics[width=3.3in]{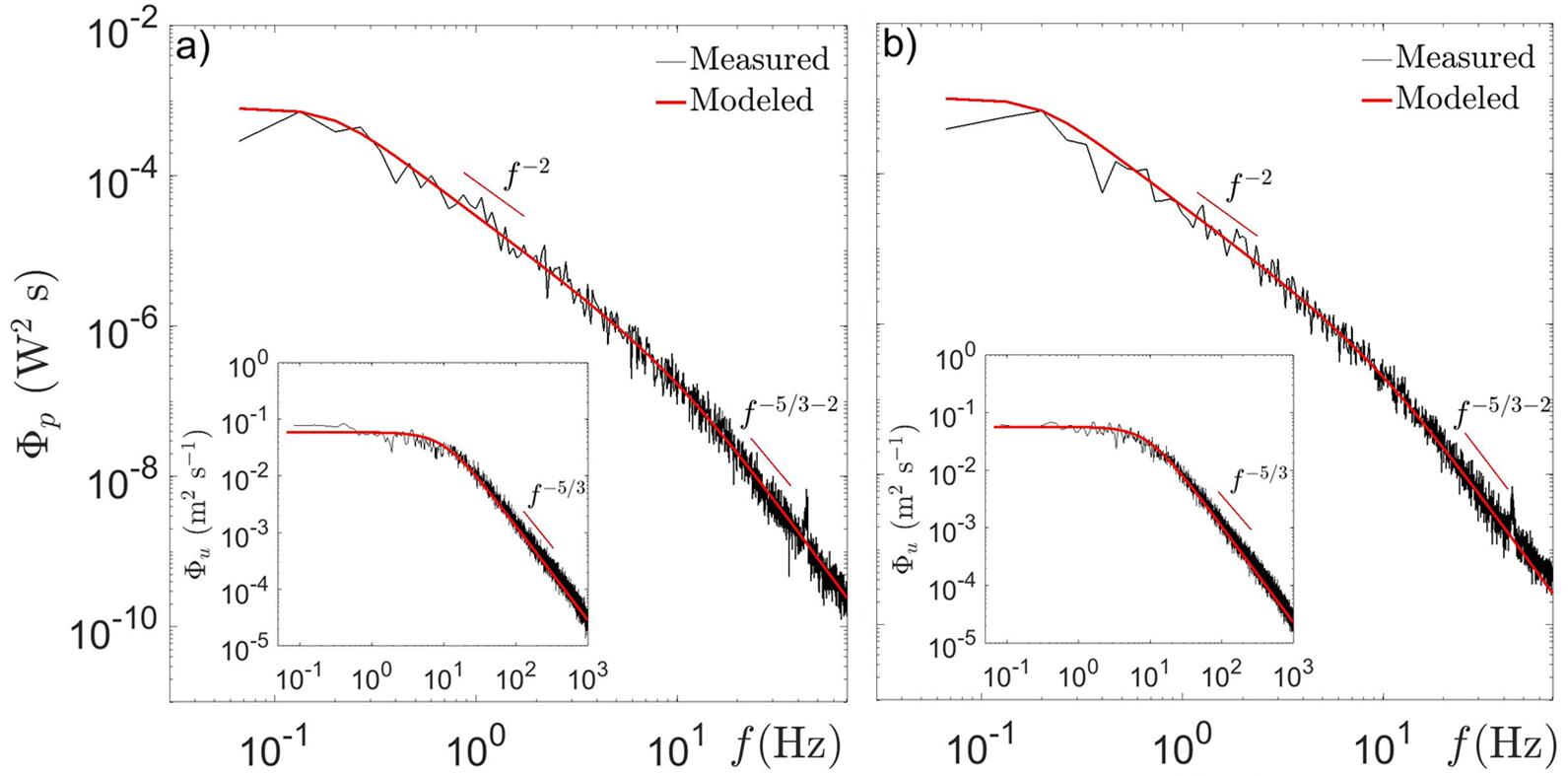}
	\caption{Measured and modeled spectra of hub-height velocity of incoming flow (sub-figures) and output power inside wind farm. a) 5th row, $S_x=7$; b) 4th row, $S_x=10$.}
	\label{datamodel}
\end{figure}

Based on the features of $\Phi_P$ from single turbines within the wind farm, we model the power fluctuations in the $i$th row in the same way as the single turbine considering the local flow at hub height. Further,  the local incoming $\Phi_u$ can be estimated with the von K\'arm\'an \citep{Von48} model spectrum ($\Phi_{u}^K$) using the local integral length scale ($\Lambda^u$) and velocity variance ($\sigma_u^2$), as follows:
\begin{equation}
	f\Phi_{u}^K(f)/\sigma_u^2=4n_u/(1+70.8{n_u}^2)^{5/6}
\end{equation}
where $n_u=f\Lambda^u/U$. In this context, $\Lambda^u$ and $U$ are representative of the incoming local flow ($i$-th row) at hub height. This procedure is shown in Figure \ref{datamodel} for the turbines in the 4th and 5th rows of the $S_x$=7 and 10. This suggests that $\Phi_{u}^K$ for the local velocity is able to properly infer the local $\Phi_u$.

\begin{figure}
	[h!]
	\centering
	\includegraphics[width=2.4in]{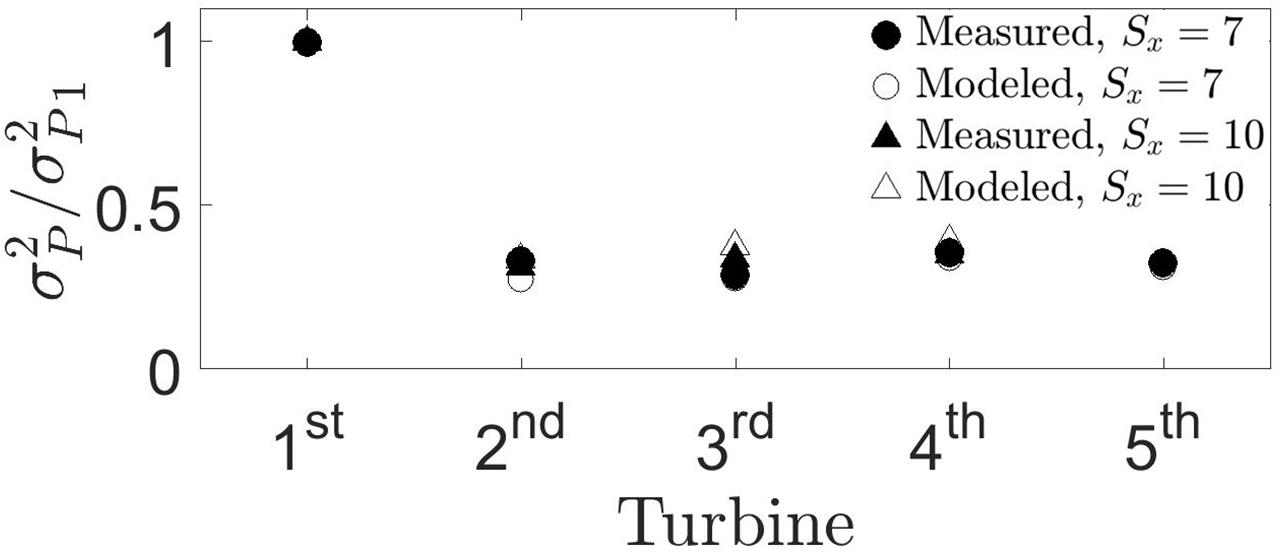}
	\caption{Measured and modeled power variance of individual turbines in the wind farm with $S_x = 7$ ($o$) and $S_x = 10$ ($\triangle$).}
	\label{powervari}
\end{figure}

Using field measurements, Morfiadakis \citep{Morfiadakis96} proposed that $\Phi_{u}^K$ is suitable for canonical boundary layers. According to Figure \ref{datamodel}, the local velocity spectrum at hub height appears to be well modeled by $\Phi_{u}^K$. This suggests that it is appropriate in regions where tip vortices have no strong effect on the flow \citep{Chamorro10}. Appropriate estimation for $\Lambda^u$, $U$ and $\sigma_u^2$ is key to allowing for the use of $\Phi_{u}^K$.
Like the case of a single turbine \citep{Tobin15b}, the filtering effect of the turbine on the power output is estimated with a second-order Butterworth filter; the cutoff frequency is the inverse of the inertial timescale, and the forward gain can be estimated by taking the velocity derivative of the turbine power equation. The resulting spectral relation is then: 
\begin{equation}
	\Phi_P(f)=\dfrac{3/2C_P\rho AU^2}{\sqrt{1+\left( 2\pi t_i\right) ^4}}\dfrac{4\sigma_u^2T^u}{\left( 1+70.8\left( fT^u\right) ^2\right) ^{5/6}}
\end{equation}
A comparison between the modeled and measured power output spectra of selected wind turbines in the 4th and 5th rows of the two layouts is given in Figure \ref{datamodel}. The modeled spectra show remarkable agreement with the power measurements and motivate the use for the collective $\Phi_P$ at wind-farm scale. Note that the spectral distribution for the two configurations clearly shows the $f^{-5/3-2}$ and $f^{-2}$ power law decays.
To assess the bulk performance of the model, a comparison between measured and modeled power variance $\sigma_P^2$ is shown in Figure \ref{powervari}. Note that the model only considers hub-height velocity. 
\vspace{-4mm}

\subsection{Wind-farm power from global incoming flow}
\begin{figure}[h!]
	\centering
	\includegraphics[width=3.42in]{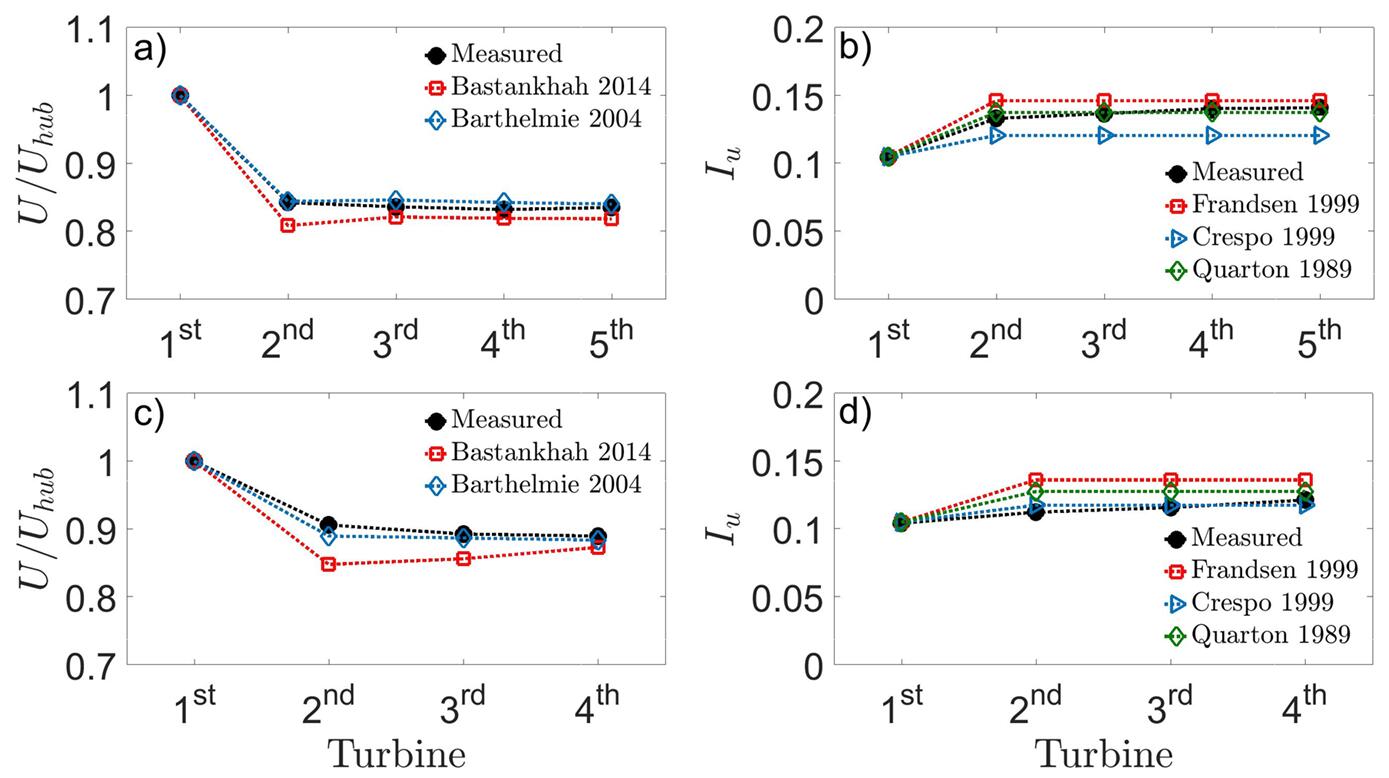}
	\caption{a) Mean velocity ($U/U_{hub}$) and b) turbulence intensity $I_u$ within the $S_x = 7$, c) Mean velocity ($U/U_{hub}$) and d) turbulence intensity $I_u$ within the $S_x = 10$ model wind farm at hub height.}
	\label{u_model}
\end{figure}

Usually, information on the incoming flow at each turbine is fairly limited. However, velocity data from the global incoming flow is likely available. Therefore, estimating the wind-farm power fluctuations with single-point measurements, namely the incoming flow at hub height of the first turbine, is very useful. Analytical models have been proposed to estimate wake flow in single turbines (e.g., \citep{Jensen83,Frandsen06,Barthelmie04,Bastankhah14}) and within wind farms (e.g., \citep{Lissaman79,Crespo99,Voutsinas90}). Another key factor is $I_u$; various formulations exists for single turbine wakes \citep{Quarton89,Hassan93,Larsen96,Crespo96,Chu14} and within wind farms e.g., \citep{Frandsen99,Niayifar15}. A comparison of the mean flow and $I_u$ for various models with the measurements is shown in figure \ref{u_model}.
It is possible to assume minor variations past 2-3 rows of turbines for practical purposes. Then, we can use the formulations for $U$ and $I_u$ to account for the local incoming flow. Here, we use the model by Voutsinas \citep{Voutsinas90} with wake velocity models to simulate the velocity distribution inside the two model wind farms. The velocity model by Barthelmie \citep{Barthelmie04} and $I_u$ model of Quarton \citep{Quarton89} are used to estimate the input parameters for the power fluctuations. Limited literature exists for $\Lambda^u$ in turbine wakes or inside of wind farms. Experiments by Chamorro \textit{et al}. \citep{Chamorro13} were found to fit well in the wind tunnel measurements by Jin \textit{et al} \citep{Jin16}. Hereon, despite some deviations with our measurements, the evolution curve from these sources was used for $\Lambda^u$.

\subsection{\textbf{Covariance due to advection and turbulent diffusion}}
\begin{figure}
 	[h!]
 	\centering
 	\includegraphics[width=3.4in]{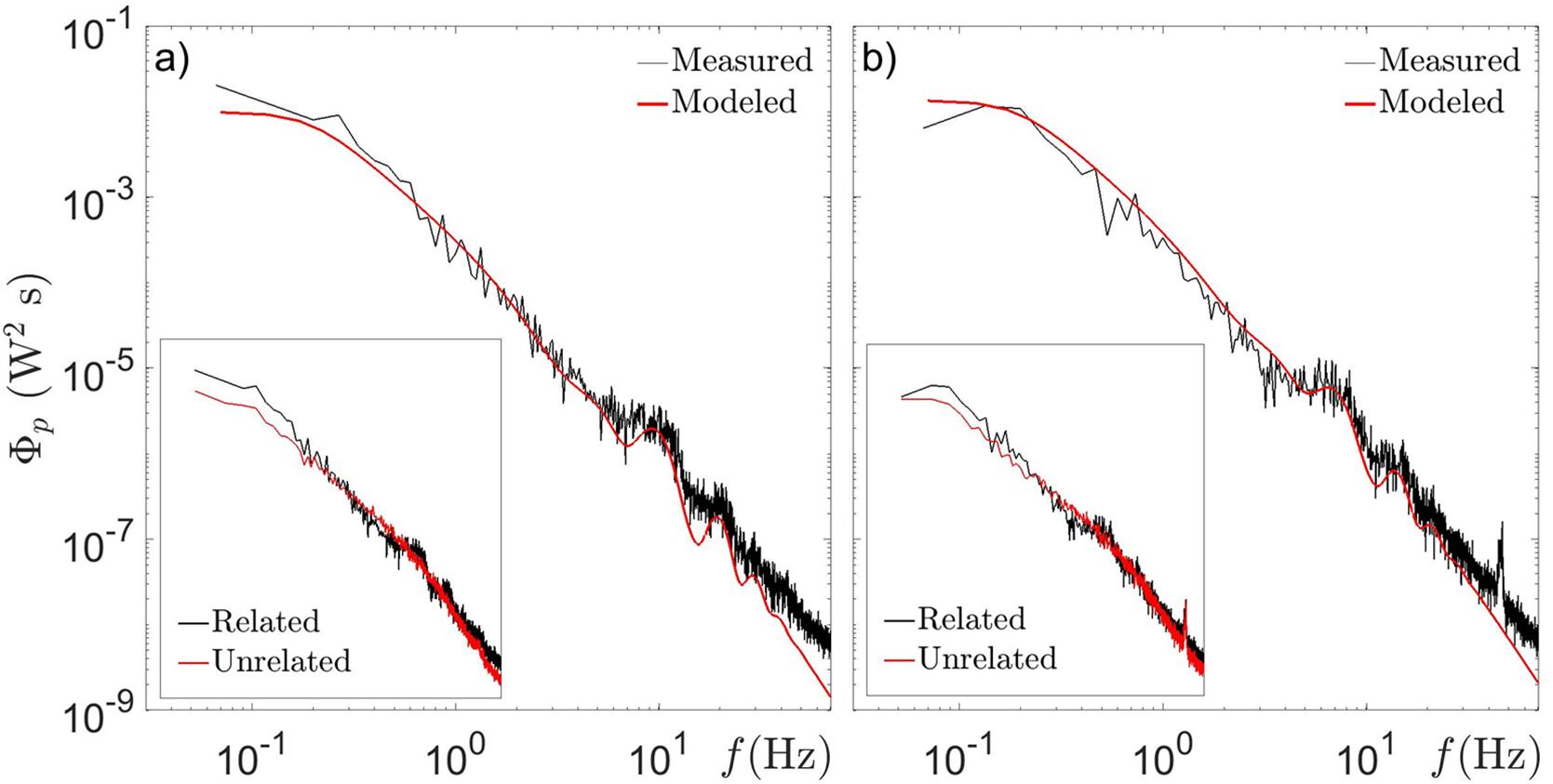}
 	\caption{Measured and modeled spectra of output power in the wind farm. a) five rows, $S_x = 7$; b) four rows, $S_x = 10$.}
 	\label{formulation_model}
 \end{figure}

Because nearby turbines simultaneously respond to large-scale atmospheric motions, the covariance of turbine pairs needs to be considered when predicting the total variance, as indicated in Equation \ref{wind_farm_power}.
\begin{equation}
	{\sigma^2\left(\sum_{i=1}^{N}{P_i}\right)}=\sum_{i=1}^{N}{{\sigma^2_{P_i}}}+2{\sum_{i=1}^{N}}\sum\limits_{j>i}^{N}{Cov\left( P_i,P_j\right) } 
	\label{wind_farm_power}
\end{equation}
The effect of covariance between turbine pairs inside of a wind farm is inspected with experiments by measuring the instantaneous power of the turbines both synchronously and asynchronously. As illustrated in the (sub-)Figure \ref{formulation_model}, covariance between turbines has a notable effect on the spectrum across scales. The first significant difference is in the low-frequency region, where neglecting covariance conspicuously under-predicts the spectral density. This is attributed to the fact that eddies with scales much larger than the separation between turbines modulate all their behaviors simultaneously. Furthermore, it is noted that significant bumps (oscillations) occur in the frequency region on the order of $U/S_x$ and its harmonics. As anticipated, the frequencies where the bumps occur in $S_x = 7$ case are larger than those of the $S_x = 10$ case correspondingly as the advection time between turbines is shorter. These bumps are attributed to motions that impart their signature on an upwind turbine, are advected downwind, and then impart their signature on a downwind turbine a short time later, leading to a periodic output. Although this leads to bumps at the advection time scale and its harmonics, the higher-frequency bumps are relatively weaker, likely due to turbulent decoherence of the small-scale structures.

To predict the power fluctuations with only incoming flow, it is necessary to estimate the covariance based on physical principles. Similar to Equation \ref{wind_farm_power}, the power spectrum of the wind farm must include a contribution of twice the co-spectrum of turbine pairs. The co-spectrum is the real part of the Fourier transform ($\mathscr{F}$) of the cross-correlation of the two power signals. The auto-correlation of the combined signal consists of the cross-correlation of the 1st signal with the 2nd, and of the 2nd signal with the 1st. The $\mathscr{F}$s of these signals are complex conjugates, which justifies taking the cross-correlation contribution as twice the real part of the $\mathscr{F}$.

Based on Taylor's frozen-eddy hypothesis \citep{Taylor38} and Kraichnan's idealized random sweeping hypothesis \citep{Kraichnan67}, Wilczek and Narita \citep{Wilczek12} proposed a model to predict the two-time wavenumber co-spectrum of a laterally homogeneous turbulent shear flow. According to this model, the two-time co-spectrum is closely related to the instantaneous energy spectrum. Because power output fluctuations are driven by the turbulence, it is  reasonable to connect the cross-correlation of the output power to that of the flow. The random sweeping hypothesis states that a frozen turbulence field is advected by the velocity $U+v'$, as given in Equation \ref{kraichnan}, where $v'$ is referred to as the sweeping velocity.
\begin{equation}
\frac{\partial u(x,t)}{\partial t}+(U+v')\frac{\partial u(x,t)}{\partial x}=0
\label{kraichnan}
\end{equation}

Considering two spatially separated points $x_1$ and $x_2$, taking the $\mathscr{F}$ of equation \ref{kraichnan} and solving for the velocity $\hat{u}=\mathscr{F}(u)$, the following result is obtained:
\begin{equation}
\begin{aligned}
\hat{u}(x_2,f)=\langle \exp \left( \frac{-2\pi i f \Delta x}{U+v'}\right)\rangle\hat{u}(x_1,f),
\end{aligned}
\end{equation}
where $\langle\rangle$ denotes temporal averaging and $\Delta x = x_2$-$x_1$.

By assuming that the sweeping velocity $v'$ is much smaller than the advection velocity, a similar approach to Wilczek and Narita is taken to model the two-point frequency spectrum. This leads to a complex exponential behavior in the co-spectrum due to advection and turbulent decoherence. By assuming a Gaussian probability density function for $v'$, the following result can be obtained for the co-spectrum:
\begin{equation}
\begin{aligned}
\phi_{1,2}=\phi_{1,1}(f)\exp \left( \frac{-2\pi i f \Delta x}{U}\right)\times \\
\exp\left(\frac{-2\pi^2 f^2 \Delta x^2 \langle v'\rangle^2}{3U^4}\right)
\end{aligned}
\end{equation}
where $\phi_{1,2}$ is the cross-spectrum of points $x_1$ and $x_2$, and $\phi_{1,1}$ is the power spectrum at location $x_{1}$. Because only the real part is taken, the complex exponential is reduced to a cosine contribution. We will further assume that $\langle v'\rangle^2=\sigma_u^2$. Thus, assuming power is nearly uncorrelated between columns in the aligned layout wind farm, which is consistent with results of \citet{Stevens14} and \citet{Bossuyt17}, we derive the spectral form of the power output of an entire wind farm in the frequency domain as follows:
\begin{equation}
\begin{aligned}
	{\Phi_p}_{wf}=M  {\sum_{i=1}^N}{\Phi_p}_i+2M{\sum_{i=1}^N}\sum\limits_{j>i}^N  {\Phi_p}_j cos\left(2 \pi f\tau_{ij}\right)\times\\ \exp\left( -\dfrac{2}{3}\pi^2 f^2{\tau_{ij}}^2{I_j}^2\right) 
\end{aligned}
	\label{(power_spectra)}
\end{equation}
where $M$ and $N$ are the number of columns in the transverse and streamwise directions. The co-spectrum of turbine pairs exhibits the product of a harmonic oscillation $cos\left( \pi f\tau_{ij}\right)$, and an exponential decay $exp(-\dfrac{2}{3}\pi^2 f^2{\tau_{ij}}^2{I_j}^2)$. The cosine portion of this formulation is from pure advection of frozen turbulence from one point upwind to another downwind. The exponential decay accounts for the fact that the turbulence is not perfectly advected, and becomes distorted as it move downwind, particularly so for high-frequency motions. Here, $\tau_{ij}=(j-i){S_x}d_T/U_j$ represents the advection time between turbines $i$ and $j$, and $I_j$ denotes the local $I_u$ of turbine $j$. Thus, ${\Phi_p}_1$ is the power spectrum of the first row and is calculated with Equation from \citep{Barthelmie04} with incoming flow as input. The power spectra of turbine $i$ ($ > 1$) inside the wind farm, ${\Phi_p}_{i}$, is calculated with the modeled parameters as input. 
 
 The predicted power output spectra of the two layouts is shown in Figure \ref{formulation_model}, with only incoming flow as input to Equation \ref{(power_spectra)}. In general, the formulation shows a good fit with measurements; the location and approximate magnitude of the bumps are also well predicted. The model does a comparatively poorer job of predicting low-frequency spectral densities. This may be due to the lack of good methods for estimating $\Lambda^u$ and the assumption of laterally homogeneous flow in the sweeping hypothesis. Further, there is inherently greater uncertainty in low-frequency spectral density measurements, which can only be alleviated with greater measurement time. It should be noted that this formulation does not account for important dynamical occurrences in the wind-farm flow, such as wake meandering \cite{Foti16}.
 
\section{Summary}
This framework aims to fill outstanding gaps in the quantification of wind farm power fluctuations. With only the global incoming flow at hub height, the model is able to estimate the structure of the power fluctuations including range and level of characteristic regions as well as spectral oscillation. For a single turbine configuration, the spectral characteristics of the power fluctuation is determined via the incoming turbulence and transfer function. Spatio-temporal correlations related to the advection and turbulent diffusion of large-scale motions lead to small bumps in the spectra of power output in a wind farm.

This work has a broad impact in the scientific and engineering communities as well as industry dealing with wind-farm power fluctuations. Instead of the instantaneous measurements of flow characteristics at the vicinity of each turbine, the framework allows for the estimation of the total power fluctuations of a wind farm using $I_u$ and $\Lambda_u$ via $\Phi_{u}^K$. As a distinctive characteristic caused by the spatio-temporal correlation of the flow, the local spectral  maximum captured in our wind tunnel measurement have also been observed in field tests (fig. 6 in \citet{Calif13}) as well as numerical simulations (fig. 6 in \citet{Stevens14}), which further verifies our framework. This study also leaves open questions for future investigation. In particular, the characterization of the integral time scale distribution in turbine wakes needs further quantification. Also, the effect of complex topography, wake meandering  and layout need to be evaluated in generic conditions. We hope that the insight can provide forward-looking guidance for the power estimation of wind farms and  better schemes controlling the power output fluctuations.

\section{Acknowledgments}
This work was supported by the Department of Mechanical Science and Engineering, at the University of Illinois, as part of the start-up of L.P.C., as well as the China Scholarship Council for H.L. This material is based upon work supported by the National Science Foundation Graduate Research Fellowship Program under Grant Number DGE-1144245. This material is based upon work supported by the National Science Foundation under Grant No. ECCS-041544081.

\nocite{*}



%

\end{document}